\documentstyle[prl,epsf,amssymb,amsbsy,amstext,amsfonts,aps]{revtex}

\begin{document}

\bibliographystyle{try}

\topmargin 0.5cm

\twocolumn

\draft{}


\wideabs{
\title{Photoproduction of Vector Mesons at Large Transfer}
\author{ J.-M. Laget }
\address{ CEA/Saclay, DAPNIA/SPhN, F91191 Gif-sur-Yvette Cedex,
France}
\date{\today}
\maketitle

\begin{abstract}
At forward angles, the cross-sections of photoproduction of vector mesons ($\rho$, $\omega$, and $\phi$) are well accounted for by the exchange of the Pomeron at high energies, while contributions of $t$ channel exchange of Reggeons are significant at low energies. At large angles, the impact parameter becomes small enough to prevent their constituents to build up the exchanged Reggeons or Pomeron. Two gluon exchange appears to dominate above $-t\simeq 1$ GeV$^2$,
especially in the $\phi$ channel.
\end{abstract}

}

\narrowtext

\begin{figure}[h]
\epsfxsize=7.5 cm
\epsfysize=7. cm
\centerline{\epsffile{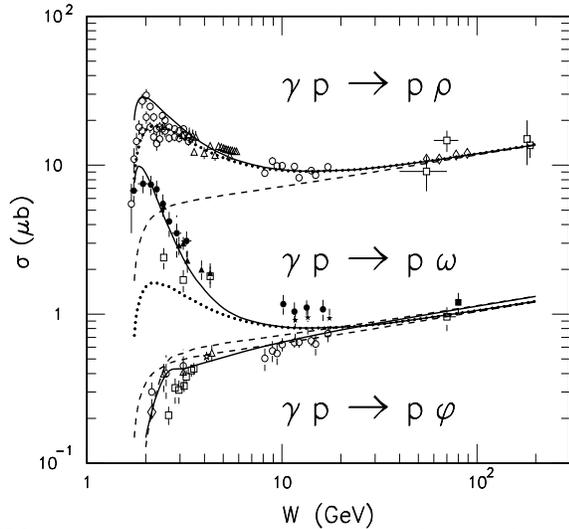}}
\caption[]{The cross-sections of elastic photoproduction of vector mesons off proton are plotted against the c.m. energy $W=\sqrt{s}$. Dashed curves: Pomeron exchange. Dotted curves: Pomeron and $f_2$ exchange. Full curves: full model. Experiments from Ref.~\cite{Dur00}}
\label{sig_tot}
\end{figure}

Elastic photoproduction of vector mesons exhibits the same slow rise with the energy as the hadronic cross-sections (Fig.~\ref{sig_tot}): The time during which the photon fluctuates into vector mesons is long enough to permit their interactions with the target nucleon.  The exchange of the Pomeron accounts fairly well for this universal behavior, while at lower energies the exchange of a few Regge trajectories is necessary to reproduce the variation of the cross-sections~\cite{La98}. At large angles, the impact parameter is small enough to prevent two gluons (resp. two quarks) to build  up the exchanged Pomeron (resp. Reggeons). This note investigates under which conditions the Pomeron and the Reggeons are resolved into their simplest constituents and how they couple to the constituent quarks in the vector meson and the nucleon.

Following the notations of Ref.~\cite{La95}, the amplitudes describing the exchange of the Pomeron and the $f_2$ meson can be combined in the compact form:
\begin{eqnarray}
 {\cal T}_P+{\cal T}_{f_2} =i \frac{12\sqrt{6}m_V e_q f_V \beta^2_0 \mu^2_0 F_1(t)}
                                           {(Q^2+m^2_V -t)(2\mu^2_0+ Q^2+ m^2_V -t)} \nonumber \\
 \times [2p\cdot q \epsilon_V \cdot \epsilon +2\epsilon_V\cdot p \epsilon \cdot (q-P_V)
             -2\epsilon_V \cdot q\epsilon \cdot p                        \nonumber \\
              +\frac{Q^2+m^2_V-t}{p\cdot q} \epsilon_V \cdot p\epsilon \cdot p]     
    \left (\left(\frac{s}{s_0}\right)^{\alpha_P (t)-1}  e^{-\frac{1}{2}i\pi \alpha_P(t)}
\right . \nonumber \\ \left . 
                 +\kappa_{f_2} \left(\frac{s}{s_1}\right)^{\alpha_{f_2} (t)-1}
                     \frac{(1+e^{-i\pi \alpha_{f_2}(t)})\pi \alpha'_{f_2}}
                            {2\sin(\pi \alpha_{f_2}(t))\Gamma (\alpha_{f_2}(t))} 
\right)     
\end{eqnarray}
being $p=(E_i,\vec{p_i})$, $q=(\omega, \vec{k})$ and $P_V=(P^0_V,\vec{P_V})$ the four momenta of the target nucleon, the incoming photon and the outgoing meson: $2p\cdot q= s-m^2+Q^2$. The Regge trajectory of the Pomeron is $\alpha_{P}(t) = 1.08 +0.25 t$, while the trajectory of the $f_2$ meson is $\alpha_{f_2}(t) = 0.55 + \alpha'_{f_2} t$, with $\alpha'_{f_2} = 0.7$ GeV$^{-2}$. The mass scales are $s_0 = 4$ GeV$^2$ and $s_1 =$ 1 GeV$^2$.

The model~\cite{Do87} assumes that the Pomeron, or the $f_2$, couples to a single constituent quark in the vector meson and in the nucleon, as a $C=+1$ isoscalar photon.  In the high energy limit, the trace over the matrices in the quark loops leads to the three first terms in the square bracket. The last term has been added to restore gauge invariance: it can be considered as a contact term.  When the strength of the coupling of the Pomeron with a quark, $\beta_0^2 \approx  4$ GeV$^2$, is fixed by  the analysis of the nucleon-nucleon high energy scattering, the Pomeron exchange amplitude reproduces the magnitude of vector meson photoproduction cross sections in the HERA energy range, $W\sim 100$ GeV. Their ratio follows from the actual values of the effective charge $e_q$ of the quark, the mass $m_V$ and the radiative decay constant $f_V$ of the meson. In the Fermi Lab, CERN  and SLAC energy range, $W\sim 10$ GeV and below, the Pomeron exchange contribution underestimates the measured cross sections. Here the exchange of a trajectory with an intercept close to $0.5$ is needed to provide a contribution which decreases slowly with energy. Since the $\rho$ or $\omega$ meson cannot be exchanged, the model includes a non-degenerate trajectory based on the $f_2(1270)$ meson, which is assumed to couple to the quark as the Pomeron, the strength of its coupling ($\kappa_{f_2}= 9$) being adjusted in order to reproduce the $\rho$ channel cross section. It reproduces also the $\omega$ channel cross section, especially its natural parity exchange component (open squares at $W=2$ and $3$ GeV, in Fig.~\ref{sig_tot}). In the $\phi$ channel, the Pomeron exchange amplitude reproduces the trend of the cross-section down to threshold, slightly overestimating it (by 20\%) at low energy. Here, the exchange of $f'_2(1525)$, which has a significant strangeness content, is preferred to the $f_2(1270)$ meson and brings the model close to the data when $\kappa_{f'_2}= -4.5$.

Scalar and pseudo scalar meson exchanges contribute at lower energies accessible at SLAC and JLab. The spatial part of the $\pi$ exchange amplitude takes the form:
\begin{eqnarray}
 {\cal T}_{\pi} = i \frac{eg_{V\pi \gamma }}{m_{\pi }}g_0 \frac{\sqrt{(E_i+m)(E_f+m)}}{2m}
        F_{\pi}(t)  \left(\frac{s}{s_1}\right)^{\alpha_{\pi }(t)}  \nonumber \\                                                                                                            
    \frac{\pi \alpha'_{\pi } e^{-i\pi \alpha_{\pi }(t)}}{\sin(\pi \alpha_{\pi }(t))\Gamma (\alpha_{\pi }(t)+1)}     
\left(\chi_f \left| \vec{\sigma }\cdot \left[ \frac{\vec{p_i}}{E_i+m}
                   \right . \right . \right . \nonumber \\ \left . \left . \left .
                                                                          -\frac{\vec{p_f}}{E_f+m}\right]\right|\chi_i\right)
\protect{
[\vec{k}\times (\epsilon^0_V \vec{P_V} - P^0_V \vec{\epsilon_V}) 
       -\omega \vec{\epsilon_V}\times\vec{P_V} ] \cdot \vec{\epsilon}
} 
\end{eqnarray}
being $(E_f,\vec{p_f})$ the four momenta of the recoiling nucleon and $\chi_i$, $\chi_f$  the nucleon spinors. The degenerate Regge trajectory~\cite{Gui97}  is $\alpha_{\pi }(t)=  (t-m^2_{\pi }) \alpha'_{\pi }$, with $\alpha'_{\pi }= 0.7$ GeV$^{-2}$. The $\pi N$ coupling constant is $g^2_0/4\pi =14.5$. According to the low energy phenomenology, a monopole form factor $F_{\pi }(t)=(\Lambda^2_{\pi }-m^2_{\pi }) /(\Lambda^2_{\pi }-t)$, with $\Lambda_{\pi}= 1.1$ GeV, takes into account the finite size of the $\pi N$ vertex.


The $\pi$ exchange amplitude dies quickly  when the energy increases, due to the nearly vanishing intercept of the $\pi$ trajectory. It contributes dominantly in the $\omega$ channel (Fig.~\ref{sig_tot}), due to the relative size of the radiative coupling constants: $g_{\omega \pi \gamma}= 0.334$, $g_{\rho \pi \gamma}= 0.136$, $g_{\phi \pi \gamma}= 0.0301$. In the $\phi$ channel, $\eta$ exchange may also be considered~\cite{WiXX}, but due to the actual size of the coupling constants it contributes at the same level as the $\pi$ exchange and is negligible.

In the $\rho$ production channel,  $\sigma$ exchange cannot be excluded near threshold. The amplitude follows the expression of Ref.~\cite{SoYY},  but with a degenerate Regge propagator. The radiative coupling constant has been adjusted in order to accommodate the highest values of the experimental cross section, but the spread of the data can also be accounted for by the $f_2$ exchange alone. 

\begin{figure}[]
\epsfxsize=7.5 cm
\centerline{\epsffile{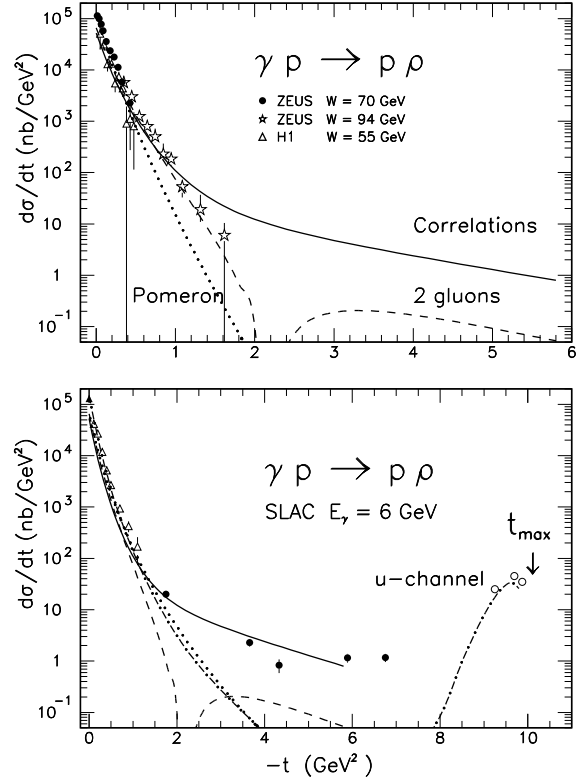}}
\caption[]{The differential cross-section of elastic photoproduction of rho mesons off proton is plotted against the four-momentum transfer $t$, in the HERA energy range (top) and JLab/SLAC energy range (bottom).}
\label{rho}
\end{figure}

It is remarkable that such a simple model reproduces over two decades the energy variation of the cross section of the three channels, at the expense of only one parameter, the relative strength of the $f_2$ effective coupling. All the other inputs (Pomeron quark or $\pi$N couplings, radiative couplings, Regge trajectories, \ldots) are determined from other channels. As illustrated in Fig.~\ref{rho}, it naturally reproduces the forward angle cross-sections as well as the shrinking of the diffractive peak when the energy increases from the SLAC~\cite{AnXY} to the HERA~\cite{Ze95} domain (dotted lines).

However, it underestimates the experiment at large transfer $t=(P_V-q)^2$: above $-t=0.5$ GeV$^2$ in the HERA energy range and $-t=1$ GeV$^2$ in the JLab/SLAC energy range. Here, the impact parameter ($b\propto 1/\sqrt{-t} \sim 0.2$ fm when $-t= 1$ GeV$^2$) becomes comparable or smaller to the gluon correlation length ($a\sim 0.2\div0.3$ fm)--- the distance over which a gluon can propagate: two gluons can be exchanged between a quark in the vector meson and a quark in the proton, without being forced to recombine into a Pomeron. Larger momentum transfers also select configurations where the transverse distance between the two quarks, which must recombine into the vector meson, is small: each gluon can also couple to a different quark. As demonstrated in Ref.~\cite{La95}, the interference between these two mechanisms leads to a characteristic node, around $-t\sim 2\div 2.5$ GeV$^2$, in the two gluon exchange cross-section (dashed lines). The color singlet nature of the $q\overline{q}$ pair imposes such a destructive interference. It is worth noting that such a cancellation is at the origin of color transparency~\cite{Gu77}: the weakening of the interaction between white objects of which the transverse size vanishes. On the same token, each gluon can also couple to a different quark inside the nucleon target, giving access to correlations between quarks in its ground state (full lines). 

Under the same assumptions, and with the same notations, as in Ref.~\cite{La95} the two gluon amplitude takes the form:
\begin{eqnarray}
{\cal T}_{2g} = \frac{6\sqrt{6}m_V e_q f_V }{8\pi } 
                       [2p\cdot q \epsilon_V \cdot \epsilon +2\epsilon_V\cdot p \epsilon \cdot (q-P_V)
                                                                                                       \nonumber \\
 -2\epsilon_V \cdot q\epsilon \cdot p +\frac{Q^2+m^2_V-t}{p\cdot q} \epsilon_V \cdot p\epsilon \cdot p]
                                                                                                       \nonumber \\
\int {\rm d}l^2 [4\pi \alpha_s D(l^2+\frac{1}{4}t)] ^2    \protect{[F_1(t)- F_1(3l^2+\frac{t}{4})]}
                                                                                                       \nonumber \\
\left( \frac{1}{Q^2+m_V^2-t} - \frac{1}{Q^2+m_V^2-4l^2} \right)
\label{2g}
\end{eqnarray}
where the integral runs over the transverse momentum $l^2$ of the exchanged gluons. It follows from the works of Refs.~\cite{La87,La89} and is the same as the expression derived in Ref.~\cite{Cu97}, except that, instead of a pertubative propagator, a dressed gluon propagator~\cite{La89} is used:
\begin{eqnarray}
\alpha_s D(l^2)=\frac{\beta_0}{\sqrt{\pi} \lambda_0}
                                             \exp\left(\frac{l^2}{\lambda_0^2}\right),
                                                 \nonumber \\
2\pi \int_ {-\infty }^{0} {\rm d}l^2 [\alpha_s D(l^2)]^2= \beta_0^2
\end{eqnarray}
The value~\cite{La95} of the range parameter $\lambda_0^2= 2.7$ GeV$^2$ corresponds to a gluon correlation length $a\approx  0.19$ fm.

The gauge invariant expression in the first square bracket results from the trace in the parton loops, assuming that the constituent quarks  of the vector meson are frozen. In Refs.~\cite{La95,La87,La89}, the two gluons were assumed to couple to the same quark of the nucleon, the structure of which being taken into account by the experimental value of the isoscalar form factor
\begin{eqnarray}
F_1(t)=\int \left[\prod {\rm d}\beta_j{\rm d}\vec{r_j}\right]
           \delta(\sum \beta_j \vec{r_j}) \delta(\sum \beta_j-1)
                  \nonumber \\
|\Psi (\beta_j,\vec{r_j})|^2   e^{i\vec{\Delta}\cdot \vec{r_k}} 
\end{eqnarray}
where the nucleon wave function $\Psi$ depends upon the transverse coordinate $\vec{r_j}$ and  the fraction of the longitudinal momentum  $\beta_j$ carried by the quark $j$, and where $\vec{\Delta}^2= t$. When each gluon is allowed to couple to a different quark, the structure of the nucleon enters eq.~\ref{2g} through the correlation function ~\cite{Gu77,Cu94}
\begin{eqnarray}
G(\vec{k_a},\vec{k_b})= \int \left[\prod {\rm d}\beta_j{\rm d}\vec{r_j}\right]
           \delta(\sum \beta_j \vec{r_j}) \delta(\sum \beta_j-1) 
                              \nonumber \\
|\Psi (\beta_j,\vec{r_j})|^2 e^{i\vec{k_a}\cdot \vec{r_k}+i\vec{k_b}\cdot \vec{r_l}}
\end{eqnarray}
Where $\vec{k_a}=\vec{l}+\frac{1}{2}\vec{\Delta }$ and $\vec{k_b}=\vec{l}-\frac{1}{2}\vec{\Delta }$ are the momenta of the exchanged gluons. Under the assumption that the gluons couple to valence quarks equally sharing the longitudinal momentum, the nucleon wave function takes the form
\begin{eqnarray}
\Psi (\beta_j,\vec{r_j})\propto \left(\prod\delta(\beta_i -\frac{1}{3})\right)
                                  \exp\left(\frac{-\sum \vec{r_i}^2}{r_N^2}\right) 
\end{eqnarray}
and the correlation function can be expressed~\cite{Gu77,Cu94,HuXY} as the isosclar form factor at a different argument
\begin{eqnarray}
G(\vec{k_a},\vec{k_b})=F_1(3l^2+\frac{t}{4})
\end{eqnarray}
Each of the two last terms in eq.~\ref{2g} corresponds to the propagator of the quark which is off-shell in the vector meson loop, when the two gluons couple either to the same quark or to two different quarks. 

Eq.~\ref{2g} exhibits in a compact way the main features of the two gluon exchange mechanisms, without free parameters. The strength, $\beta_0^2 = 4$ GeV$^2$,  of the effective coupling of the gluon to the nucleon is determined by the analysis of the nucleon-nucleon scattering, the range parameter $\lambda_0^2$ of the gluon dressed propagator corresponds to a reasonable value of the gluon correlation length, and the correlated structure of the nucleon enters through the experimental values of its form factor. It provides us with a solid starting ground for looking for such a mechanism in experiments. 

A recent experiment performed at HERA~\cite{Ze99} confirms this picture. As shown in Fig.~\ref{rho}, these data clearly favor two gluon exchange above $-t=0.5$ GeV$^2$ but, due to the statistics, cannot distinguish between the two extreme two gluon curves. They differ above $-t=1$ GeV$^2$ where  a old experiment performed at SLAC~\cite{An71}, at lower energy, confirms that the mechanism which dominates at high momentum transfer is the coupling of each gluon to two different quarks both in the vector meson and in the nucleon.

However, quark exchange cannot be excluded in the $\rho$ production channel, in the SLAC/CEBAF energy range. Their contribution has been evaluated using saturating Regge trajectories ($\alpha_M(-\infty)= -1$) in the $\pi$, $f_2$ and  $\sigma$ meson exchange amplitudes, as explained in Ref.~\cite{Gui97}. The $\pi$ exchange process leads to a contribution too small by two orders of magnitude. The $f_2$ exchange leads to a vanishing contribution, due to the signature $1+e^{-i\pi \alpha_{f_2}(t)}$ of its undegenerated trajectory. Only $\sigma$ exchange may lead to a contribution comparable to the data, but this is unconclusive due to the uncertainty on the amount of its contribution at low energy (cf. Fig.~\ref{sig_tot}).

\begin{figure}[h]
\epsfxsize=7.5 cm
\centerline{\epsffile{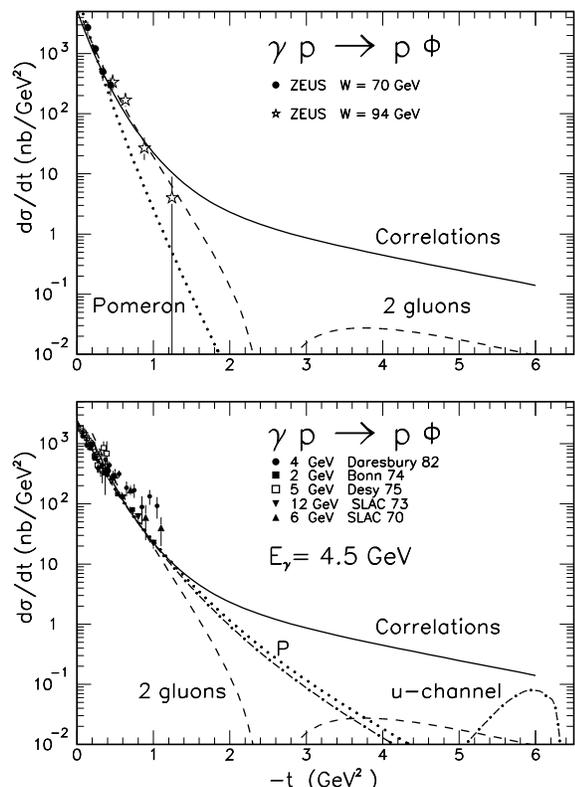}}
\caption[]{The differential cross-section of elastic photoproduction of $\phi$ mesons off proton is plotted against the four-momentum transfer $t$, in the HERA energy range (top) and JLab energy range (bottom).}
\label{phi}
\end{figure}

Due to its dominant $s\overline{s}$ component, quark exchange mechanisms are strongly suppressed in $\phi$ photoproduction, making this channel the most suitable place to look for gluon exchanges. Fig.~\ref{phi} compares model  predictions to data recently obtained at HERA~\cite{Ze95,Ze99} as well as previous data (see~\cite{La95} for references) in the JLab energy range. As in the $\rho$ channel, the Regge model reproduces the data up to $-t=0.5$ GeV$^2$ in the HERA energy range and $-t=1$ GeV$^2$ in the JLab energy range, and underestimate the data above. At high  transfer the HERA data clearly prefer the two gluon exchange model, but are unable to distinguish between its two extreme versions. This is the goal of JLab experiment 93-031, which has been recently completed~\cite{An00} in the domain $3\leq E_{\gamma}\leq 5.5$ GeV, $1\leq -t\leq 6$ GeV$^2$. It is remarkable that the preliminary results~\cite{The00} fall on the full curve in the bottom panel of Fig.~\ref{phi}. A detailed comparison with those data will be available in the experimental paper~\cite{An00}.

At backward angles (largest $t$ but smallest $u$) exchange of baryons in the u-channel may contribute in the JLab energy range. Only the exchange of the nucleon Regge trajectory is allowed in the $\phi$ (as well as $\omega$) channel. The spatial part of the amplitude is as  follows:
\begin{eqnarray}
{\cal T}_N = i\frac{e\;\mu_pg_V(1+\kappa_V)}{2m}
                  \left( \chi_f\left|\vec{\sigma}\cdot \vec{k}\times \vec{\epsilon} \;
                   \vec{\sigma}\cdot \vec{P_V}\times \vec{\epsilon_V} \right| \chi_i \right)  						\nonumber \\
    \left(\frac{s}{s_1}\right)^{\alpha_N(u)-\frac{1}{2}}                                                                                                              
    \frac{\pi \alpha'_N(1- e^{-i\pi (\alpha_N(u)+\frac{1}{2})})}
           {2\sin(\pi(\alpha_N(u)+\frac{1}{2}))\Gamma (\alpha_N(u)+\frac{1}{2})}    	       
\end{eqnarray} 
where the non degenerate nucleon trajectory is $\alpha_N(t) = -0.37 + \alpha'_N t$, with $\alpha'_N = 0.98$ GeV$^{-2}$. This amplitude leads to a good accounting of the backward angle cross-section in the $\omega$ channel~\cite{Gui97b} where all the coupling constants are fixed. In the $\phi$ production channel, the dot-dashed curve in Fig.~\ref{phi} results from the SU3 choice of the unknown $\phi$NN coupling constant $g_V(1+\kappa_V)$. In the rho channel, the $\Delta$ degenerate trajectory~\cite{Gui97b} may be exchanged: it interferes with nucleon exchange and leads to a good account  of the backward angle $\rho$ meson photoproduction (dot-dashed line in Fig.~\ref{rho}). 

In the highest part of the JLab energy range, the $u$-channel exchange backward peak and the diffractive forward peak are far enough to leave room for hunting gluon exchange processes, and studying the structure of the proton. It remains to investigate whether effects, which are negligible at higher energies, play a role: real part of the amplitude, motion of the quarks in the vector meson loop~\cite{Ro99}, full energy dependence~\cite{Va96}, etc\ldots

In summary, two gluon exchange mechanisms are good candidate to understand vector meson photoproduction at large $t$. The model which we have developed provides a good starting ground as it incorporates the basic features and its inputs are strongly constrained  by other channels.  It reproduces data recently obtained in two extreme energy domains: HERA and JLab.
It provides us with a tool to investigate quark correlations in the nucleon ground state and to test more realistic wave functions.

Discussions with M.  Diehl and M. Vanderhaeghen are greatly acknowledged.

\end{document}